# An instance-based learning approach for evaluating the perception of ride-hailing waiting time variability

*Nejc Geržinič, Oded Cats, Niels van Oort, Sascha Hoogendoorn-Lanser, Michel Bierlaire, Serge Hoogendoorn*

## Abstract

Understanding user's perception of service variability is essential to discern their overall perception of any type of (transport) service. We study the perception of waiting time variability for ride-hailing services. We carried out a stated preference survey in August 2021, yielding 936 valid responses. The respondents were faced with static pre-trip information on the expected waiting time, followed by the actually experienced waiting time for their selected alternative. We analyse this data by means of an instance-based learning (IBL) approach to evaluate how individuals respond to service performance variation and how this impacts their future decisions. Different novel specifications of memory fading, captured by the IBL approach, are tested to uncover which describes the user behaviour best. Additionally, existing and new specification of inertia (habit) are tested. Our model outcomes reveal that the perception of unexpected waiting time is within the expected range of 2-3 times the value-of-time. Travellers seem to place a higher reward on an early departure compared to a penalty for a late departure of equal magnitude. A cancelled service, after having made a booking, results in significant disutility for the passenger and a strong motivation to shift to a different provider. Considering memory decay, our results show that the most recent experience is by far the most relevant for the next decision, with memories fading quickly in importance. The role of inertia seems to gain importance with each additional consecutive choice for the same option, but then resetting back to zero following a shift in behaviour.

## 1 Introduction

In recent years, ride-hailing services like Uber, Lyft, DiDi etc. have become commonplace around the world and have thus garnered significant attention within the scientific community. Several mode choice studies, predominantly with a stated preference (SP) approach, aimed to evaluate their role and potential within the wider mobility ecosystem (Choudhury et al., 2018; Frei et al., 2017; Geržinič et al., 2022; Y. Liu et al., 2018; Yan et al., 2019). Of particular interest is the relation with public transport, as ride-hailing seems to be abstracting most of its passengers from mass transit (Henao & Marshall, 2018). Whether they compete with each other for passengers or complement each other is still subject to discussion and seems to be quite context-specific (Cats et al., 2022; Erhardt et al., 2022; Phun et al., 2019; Tirachini & del Río, 2019).

In terms of their service characteristics, ride-hailing (and other on-demand services like micro-transit and taxis) shares similarities with public transport (i.e. travellers do not need to own a car and drive themselves) as well as with a private car (i.e. offering a high level of flexibility, not being bound by spatio-temporal operations). Services do not operate according to a predefined schedule or route and may therefore be prone to a certain level uncertainty, most notably in the variability of waiting and in-vehicle time. Understanding how users perceive this variability is crucial for the future implementation of ride-hailing and its potential integration with fixed public transport.

Despite the inherent uncertainty associated with on-demand services, little is known about the impact of their service variability on mode choice. To the best of our knowledge, Bansal et al. (2019) and Alonso-González et al. (2020) were the only ones to study the effect of waiting time variability of ride-hailing services. The former performed a stated preference experiment wherein they presented respondents with two unlabelled ride-hailing services, each with its own estimated waiting time and an average pick-up delay. One service was fully reliable (average pick-up delay of 0 minutes) but had a longer expected waiting time. They report that the more unreliable the service is, the higher the percentile of the displayed waiting time should be, to avoid overly optimistic waiting time estimates with large potential pick-up delays. In a second experiment, they presented respondents with a hypothetical past experience, given a certain wait time and pick-up delay, and asked the respondents if they would switch to a different company after this experiment or not. In their sample, the majority chose to switch, with a binary model showing the propensity to switch increasing with the relative pick-up delay (delay/waiting time). Alonso-González et al. (2020) carried out three separate SP experiments, investigating the perception of variability in travel time, waiting time at the origin and waiting time at a transfer point. Similar to Bansal et al. (2019), the reliability was stated explicitly, but instead of giving respondents an average delay, three equally likely travel/wait times were presented. This gave respondents more insight into the variance of a service. Using a latent class approach, they identified four distinct groups in the population, although they differed mainly in their time valuation, while the reliability perception was similar across the segments. For all three experiments, the value of reliability was found to be lower than that of the planned waiting/travel time, with the ratio between time and reliability at around 0.5.

In contrast to the ride-hailing services context, the perception of travel time variability has been heavily researched for other modes of transport (Noland & Polak, 2002). Measuring the perception of variability through stated preference (SP) experiments is challenging, and two overarching approaches have emerged in literature; providing the information **explicitly** or **implicitly**. Presenting reliability information explicitly means that respondents are given information on the variability of potential outcomes upfront, before making a choice. Variability is therefore an attribute, that can be presented in different ways: providing a mean and standard error, an average value, a certain number of equally-likely-to-occur values or images of distributions. Both aforementioned studies on the reliability perception for ride-hailing services (Alonso-González et al., 2020; Bansal et al., 2019) apply this approach.

Alternatively, the information on variability can be given implicitly, with the respondents receiving an indication of the expected level of service and then receiving feedback on the outcome, i.e. how their selected alternative performed. By repeating this choice task several times, respondents get a more realistic experience of variability, and can thereby internalise it and form their own assessment of the reliability associated with each alternative. These experiments can be differentiated based on:

1. **Level of information provided** to the respondent, presenting only the outcome of the selected alternative or of all options (Bogers et al., 2007)
2. **Addition of a "memory aid"** for the respondent, showing them the last few outcomes (Bogers et al., 2007)
3. **Incentives / Penalties** for respondents, linked to the performance of their choices (e.g. making them wait proportionally longer if they chose an alternative with a comparatively poorer outcome (Bogers et al., 2007) or offering financial compensation based on their performance (Ben-Elia et al., 2013) in order to make the experiment more realistic). Studies may include this because there is no real "suffering" from making poor decisions in an SP experiment, and the respondents might thus not be incentivised to make choices in the same way.
4. **Number of repetitions** of such a choice task (varying from 25 (Bogers et al., 2007) up to 120 (Yu & Gao, 2019))

This type of experiment has so far predominantly been applied to car-route choice (Avineri & Prashker, 2006; Ben-Elia et al., 2008, 2013; Ben-Elia & Shiftan, 2010; Bogers et al., 2007; Tang et al., 2017; Yu & Gao, 2019) and to the best of the authors' knowledge, has not yet been used to analyse ride-hailing waiting time.

When taking part in such a survey, respondents learn the reliability of each alternative through experience. It is therefore important to consider the different types of information provided to and used by the respondents in the decision-making process. According to Ben-Elia & Avineri (2015), there are three types of information that respondents may make use of:

1. **Descriptive**: Providing the respondents with information on the state of the system
2. **Prescriptive**: Giving advice to respondents on how to decide
3. **Experiential**: The respondent's own information from the outcomes of previous choices

While descriptive and prescriptive information tends to be choice-set-specific, relevant at the time of the decision-making, experiential information is gathered over a longer period of time and has the potential to influence future decisions. Experience tends however to fade over time (Daneman & Carpenter, 1980), with more recent experiences having a stronger presence in our minds and thus a stronger influence on our decision-making (Anderson et al., 2004). Decay functions are commonly used to capture memory fading, with the most prominent example being the power function (Anderson et al., 2004; Kahana & Adler, 2002). The power function based on the work of Anderson et al. (2004) was adapted for the field of transportation by Tang et al. (2017) and applied in the context of car route choice.

There is evidence to suggest that an additional and potentially contradictory effect might also inter-play, namely human tendency to form and stick to habits, as well as our resistance to change (Gao & Sun, 2018). Many choices we make in our everyday life are habitual and not every situation is thoroughly evaluated, purely to conserve mental effort. When faced with a similar choice situation several times, people will often forego the mental effort of re-evaluating the alternatives in detail and simply keep with the alternative they are more used to. While this repeated selection of the same option can start because of better performance, people are likely to stick to it after a while, even when it is less advantageous, purely because of their habitual behaviour or inertia (Cherchi & Manca, 2011; Gao et al., 2021; Gao & Sun, 2018; González et al., 2017; Ramadurai & Srinivasan, 2006; Rashedi et al., 2017). Choice inertia is the additional value travellers assign to an alternative solely because they are used to it and the perceived mental effort required to re-evaluate the available options exceeds the perceived benefit that could be gained from a thorough analysis thereof. There is a certain threshold, above which people will still re-evaluate their choices and potentially switch, but the benefit needs to be worthwhile for the mental effort to be induced.

The goals of this study are twofold. Firstly, we identify the value travellers place on unexpected outcomes of ride-hailing trips. We focus on the variability of waiting time, as it tends to be perceived more negatively than in-vehicle time (Wardman, 2004), meaning its variability is likely to have a stronger impact on the behaviour of travellers. We also include the possibility of cancelled trips in our study, as they may have a profoundly important impact on respondents future choices. Understanding travellers' behaviour is important for operators, authorities and policymakers to know how to design such services and which aspects to prioritise. Secondly, related to the first goal, our ambition is to evaluate existing and test novel theoretical formulations of both memory decay and choice inertia, to determine how best to explain traveller behaviour in our survey, as well as in future studies on experiential learning.

The remainder of the paper is structured as follows: the survey design, model estimation and data collection are outlined in Sections 2.1, 2.2 and 2.3 respectively. The key findings are presented in Section 3, followed by a discussion of those findings and outlooks for future research in Section 4.

## 2 Methodology

### 2.1 Survey design

To evaluate the perception of waiting time variability for ride-hailing services, behavioural data of travellers is required. Our study utilises an SP experiment, as (1) revealed preference (RP) data for ride-hailing is scarce and (2) the applied model estimation approach requires a panel structure of the data (multiple observations per individual), which may be difficult to procure through RP data. We employ a survey design similar to what has been applied by Ben-Elia et al. (2013) and Bogers et al. (2007). The survey structure is presented in a flowchart in Figure 1, and described below.

Respondents are presented with two alternatives, with descriptive information on two independent ride-hailing companies; referred to as Company A and Company B to avoid any bias. The descriptive information includes **expected waiting time**, **travel time** and **trip cost**. Travel time is included for context purposes and is equal for both companies (20 min) to ensure that the trade-off behaviour is limited to the cost and waiting time aspects. Respondents are asked to choose one of the companies and are then shown their experienced waiting time. The experienced waiting times are randomly drawn from an underlying distribution, specific to each company. This process is repeated for 32 instances.

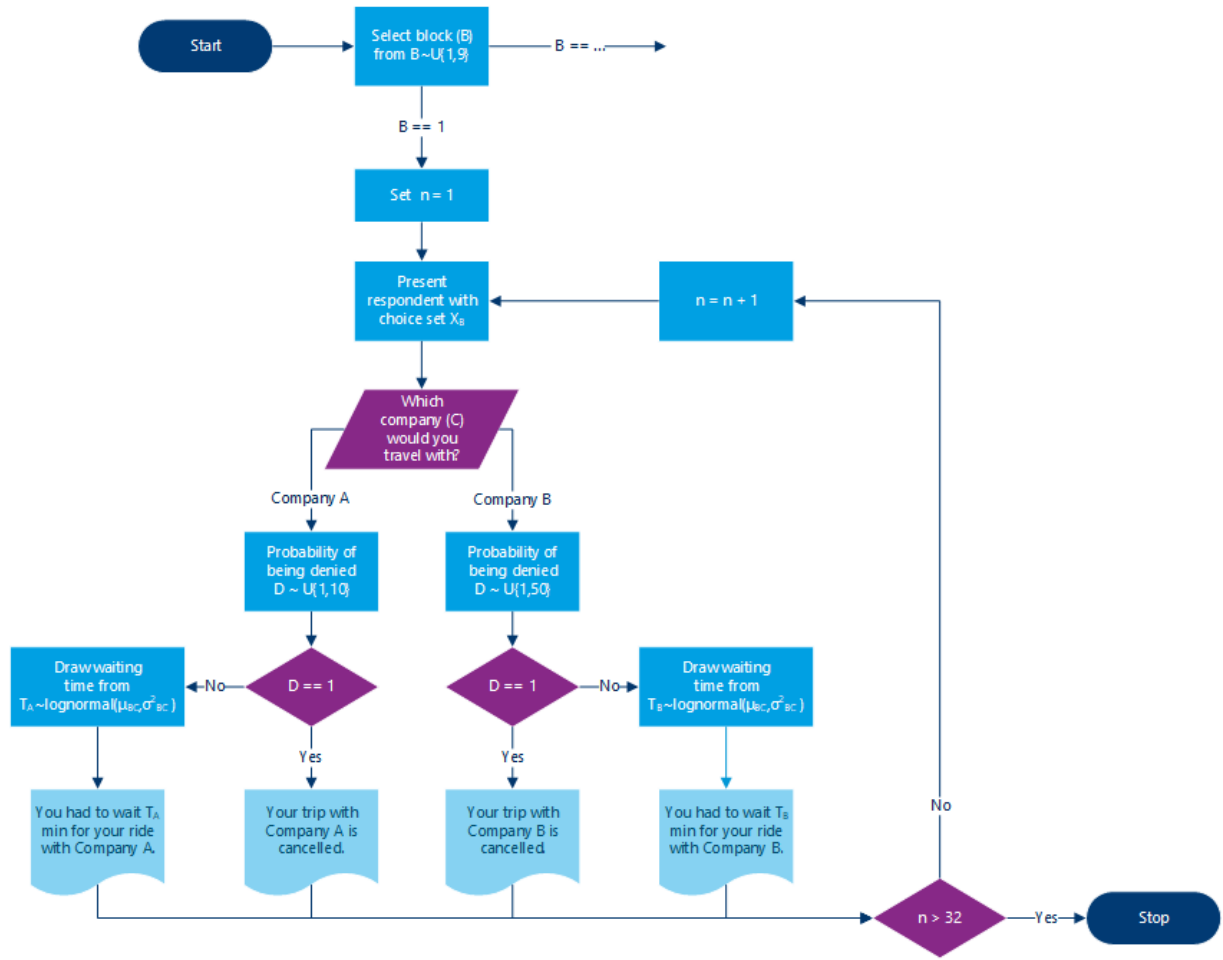

*Figure 1. Flowchart of the SP experiment focused on experiential learning*

Experiential learning experiments range in length from 25 (Bogers et al., 2007) to 120 repetitions (Yu & Gao, 2019), to ensure that the experiment is sufficiently long for the respondents to learn the reliability of each alternative. In this study, great care was also taken to make sure that the survey is not too long, to avoid overwhelming or exhausting the respondents. A pilot survey with 50 repetitions was carried out. At the start and after every ten repetitions, the respondents were asked a series of fatigue-related questions. Given the outcomes and comments from the respondents of the pilot, we limit the survey to 32 repetitions.

For the underlying distributions, a log-normal distribution is chosen as the most appropriate (Alonso-González et al., 2020; Cats et al., 2022; T. L. K. Liu et al., 2019). To add sufficient variability to the experiment, we specify three log-normal distributions and three different price levels, resulting in a total of nine combinations (Table 1). In all cases, Company A is the cheaper, but more variable alternative, while Company B is more expensive but more reliable. To determine the parameters of the log-normal distribution and the price levels, we use the results reported by Alonso-González et al. (2020), as their work is both recent and carried out in the Dutch context, making it suitable for our study. We specify the mean and variance for the log-normal distributions, which are then constructed using the µ and σ parameters (see Equation 1 and Equation 2). The mean of all distributions is set to 10 minutes, to focus on waiting time variation. The descriptive information on the expected waiting time is based on the median of each distribution. As the medians differ between the three utilised distributions, this enables us to test the also the perception of expected waiting time against the variation from it.

*Equation 1. Definition of μ parameter*

$$\mu = \ln\left(\frac{mean}{\sqrt{var + mean^2}}\right)$$

*Equation 2. Definition of σ parameter*

$$\sigma = \ln\left(\frac{var}{mean^2} + 1\right)$$

*Table 1. Log-normal distribution parameters and price levels for Company A and Company B*

| mean = 10 | $var_A$: 40 $var_B$: 10 *median: 8* | $var_A$: 40 $var_B$: 1 *median: 9* | $var_A$: 10 $var_B$: 1 *median: 10* |
|---|---|---|---|
| $Cost_A$: €12.00 $Cost_B$: €13.50 | **Block 1** | **Block 4** | **Block 7** |
| $Cost_A$: €12.00 $Cost_B$: €15.00 | **Block 2** | **Block 5** | **Block 8** |
| $Cost_A$: €13.50 $Cost_B$: €15.00 | **Block 3** | **Block 6** | **Block 9** |

In addition to the waiting time variability, we also study passengers' reaction to cancelling the ride-hailing trip altogether. The probability of the trip being cancelled is handled separately from the waiting time. Both companies are associated with a fixed cancellation probability, irrespective of the blocks, namely a 10% cancellation probability for Company A and a 2% cancellation probability for Company B. This way, service cancellation is decoupled from the waiting time distribution, while keeping with the notion of an overall more (B) and less (A) reliable company. Respondents are told of the possibility of having the trip cancelled at the start of the survey, but no additional information is provided. If the trip is cancelled, they are informed of this with a message saying that "*the trip could not be performed by your selected company*".

Additionally, respondents are presented with certain context information for the choice at the start of the survey. The trip is presented as returning home from a leisure activity (restaurant, cinema, visiting family/friends) in the evening. It is decided to avoid providing additional information on the waiting location and rather ask respondents at the end of the survey how they imagined the situation while answering the survey: was that indoors or outdoors and sitting or standing. A fifth option of "Did not really think about it" is also added. Furthermore, to compare the outcomes of the survey with the direct opinion of the respondents, they are asked which company they found more reliable. For the event of a cancelled service, we ask the respondents how they would have gotten home in real life and how their behaviour with respect to these services would change in the future. Finally, to evaluate if experience has a role in behaviour, we present the respondents with different types of services from the sharing economy (ride-hailing, car sharing, food delivery, home rental,…) and ask them whether they are aware of them and how often they make use of them. The full list of questions and possible answers is shown in Table 6 in the Appendix.

## 2.2 Model estimation

The specific nature of experiential learning data means it violates one of the principal assumptions of MNL discrete choice models (Ben-Elia & Shiftan, 2010): the error terms are not identically and independently distributed (i.i.d.). This is because the observations are interdependent and thus correlated. To circumvent this issue, we apply the Panel Mixed Logit (ML) model on our data, which relaxes the i.i.d. assumption by accounting for the panel structure of the data. Using an ML model also allows us to capture respondent heterogeneity by allowing parameters to vary between respondents.

The structure of the utility function is based on the design of the survey and how respondents were presented with the alternatives. The function consists of three main parts: (1) descriptive information, (2) experiential information and (3) the error term (shown in Equation 3). Descriptive information is formulated following the linear-additive utility. We consider the attributes that are presented to the respondents before making their choice. These include the estimated waiting time and travel cost. Travel time is not included as it is a context variable and thus equal for all alternatives (Equation 4). The error terms are assumed to be i.i.d. extreme value across all dimensions.

*Equation 3. Specification of the systematic utility*

$$U_{n,j,t} = D_{n,j,t} + E_{n,j,t} + \varepsilon_{n,j,t}$$

where:
- $U_{n,j,t}$      Total utility observed by respondent *n* for alternative *j* in choice situation *t*
- $D_{n,j,t}$      Descriptive information for respondent *n* for alternative *j* in choice situation *t*
- $E_{n,j,t}$      Experiential information for respondent *n* for alternative *j* in choice situation *t*
- $\varepsilon_{n,j,t}$      Error term associated with respondent *n* for alternative *j* in choice situation *t*

Experiential information is accumulated as respondents gain experience and is based on the feedback they receive upon making their choice (Equation 5). The two direct experiences from the survey are actual waiting time and cancelled service. The latter is coded as a binary variable, where 0 means the travel service was performed and 1 means that the trip was cancelled. The experience of waiting time is coded as the difference from the expected waiting time, meaning it can take a negative (shorter than expected waiting time) or a positive value (longer waiting time).

*Equation 4. Specification of the descriptive information part of the utility*

$$D_{n,j,t} = ASC_j + \beta_c \cdot c_{n,j,t} + \beta_{\bar{w}} \cdot \bar{w}_{n,j,t}$$

where:
- $ASC_j$      Alternative specific constant, associated with alternative *j*
- $\beta_c$      Taste parameter associated with attribute c
- $c_{n,j,t}$      Cost experience by respondent *n*, for alternative *j* in choice situation *t*
- $\bar{w}_{n,j,t}$      Estimated waiting time

In the survey, respondents make repeated choices over a period of time and the number of experiences increases. According to the instance-based learning theory (IBLT), the more recent and more frequent instances / experiences are more prominent in one's memory (Gonzalez et al., 2003; Gonzalez & Lebiere, 2005). In other words, the importance of an experience decays over time. Mathematically, this memory decay can be captured by a decay function (Equation 6), specifically the power decay function is often cited as a good approximation for modelling memory decay (Kahana & Adler, 2002; Tang et al., 2017; Yu & Gao, 2019). By incorporating the decay function into the utility function and parameterising it, it is possible to estimate the rate of memory decay for a particular situation. Most theories (including IBLT) consider the most recent experience to be the most important for the next choice situation.

*Equation 5. Specification of the experiential information part of the utility function*

$$E_{n,j,t} = \beta_w \cdot \sum_{t'}^{H_{n,j,t}} \left( (w_{n,j,t'} - \overline{w}_{n,j,t'}) \cdot m_w(t,t') \right) + \beta_x \cdot \sum_{t'}^{H_{n,j,t}} \left( x_{n,j,t'} \cdot m_x(t,t') \right) +$$

$$+ \beta_b \cdot b_{n,j,t} + \beta_i \cdot i_{n,j,t}$$

where:

| | |
|---|---|
| $w_{n,j,t'}$ | Waiting time experienced by respondent *n* for alternative *j* in experience *t'* |
| $\overline{w}_{n,j,t'}$ | Expected waiting time for respondent *n* in alternative *j* and experience *t'* |
| $x_{n,j,t'}$ | Binary variable indicating a cancelled service, experienced by respondent *n* for alternative *j* in experience *t'* |
| $m_w(t,t')$ | The weight related to memory decay of attribute *w*, in choice situation *t* for experience *t'* |
| $H_{n,j,t}$ | The set of accumulated experiences, obtained by respondent *n* for alternative *j*, when making the decision in choice situation *t* |
| $b_{n,j,t}$ | Binary variable indicating the barrier to entry, for respondent *n*, alternative *j* in choice situation *t* |
| $i_{n,j,t}$ | Binary variable indicating the inertia (habit), for respondent *n*, alternative *j* in choice situation *t* |

When modelling the impact of multiple instances on the current choice, the mathematical formulation used by Tang et al. (2017) employs an averaging approach. In other words, each individual weight of an experience is divided by the sum of all weights associated with the specific alternative. This means that the sum of all weights applied to the experiences associated with a specific alternative (not accounting for the actual utility contribution) is always equal to one (Equation 6). We refer to this as the "Relative weights" formulation. In addition, we propose and test the "Absolute weights" formulation. The notion behind this is that having more equally bad experiences should weigh more negatively and thus result in a higher disutility than a single bad experience of the same magnitude. Applying the relative weights approach, the contribution of both is equal. We amend this by removing the denominator from Equation 8, resulting in the mathematical formulation presented in Equation 7. This straightforward change means that while (for the same value of $d_z$) the ratios between weights remain the same, their total contribution to the utility changes.

*Equation 6. Specification of memory decay with relative weights of experiences*

$$m_z(t,t') = \frac{(t-t')^{-d_z}}{\sum_{t' \in H_{n,j,t}} (t-t')^{-d_z}}$$

where:
  $d_z$   Memory decay parameter associated with attribute *z*

*Equation 7. Specification of memory decay with absolute weights of experiences*

$$m_z(t,t') = (t-t')^{-d_z}$$

In addition to testing relative and absolute weights of experiences, we also test how experiences may be 'stored' in a decision-maker's memory. In the formulation of Tang et al. (2017), an experience is linked to the moment in time when it is obtained. This means that the weight associated with a given experience decreases over time, regardless of any new experiences obtaining afterwards. We refer to this as the "Time-based" formulation. To that end, we propose an "Event-based" formulation for the set of gained experiences. The notion behind this is that the most recent experience may stay at the top of the decision-maker's mind, largely irrespective of how long ago it happened. In this formulation, the only thing that may decrease the weight of an experience is obtaining a new, more recent experience with the same company. To accommodate this, the formulation of memory decay weight is adapted as indicated in Equation 8. The main difference is decoupling the power function from a time-based approach and associating it with the number of experiences *t* in a ranked order of experiences with a specific alternative. For numerical reasons, one is added to the equation, so that the most recent experience is associated with an absolute weight of 1.

*Equation 8. Specification of memory decay with an Event-based ordering of experiences*

$$m_z(t, t') = \left(|H_{n,j,t}| + 1 - h'\right)^{-d_n}$$

where:

| | |
|---|---|
| $\|H_{n,j,t}\|$ | Total number of experiences of decision-maker *n*, with alternative *j* when making the decision in choice set *t* |
| h' | The order number of the experience (1 if it is the first experience, 2 if it is the second,...) obtained at time t' |

To highlight the four different approaches and how the weights differ amongst them, they and their relative importance for the overall utility contribution are presented in Figure 2. Note that this assumes the alternative to which these weights apply was chosen at instances t=1 and t=3, therefore their impact on the choice only appears in the first following choice task. The weights are based on d=1.

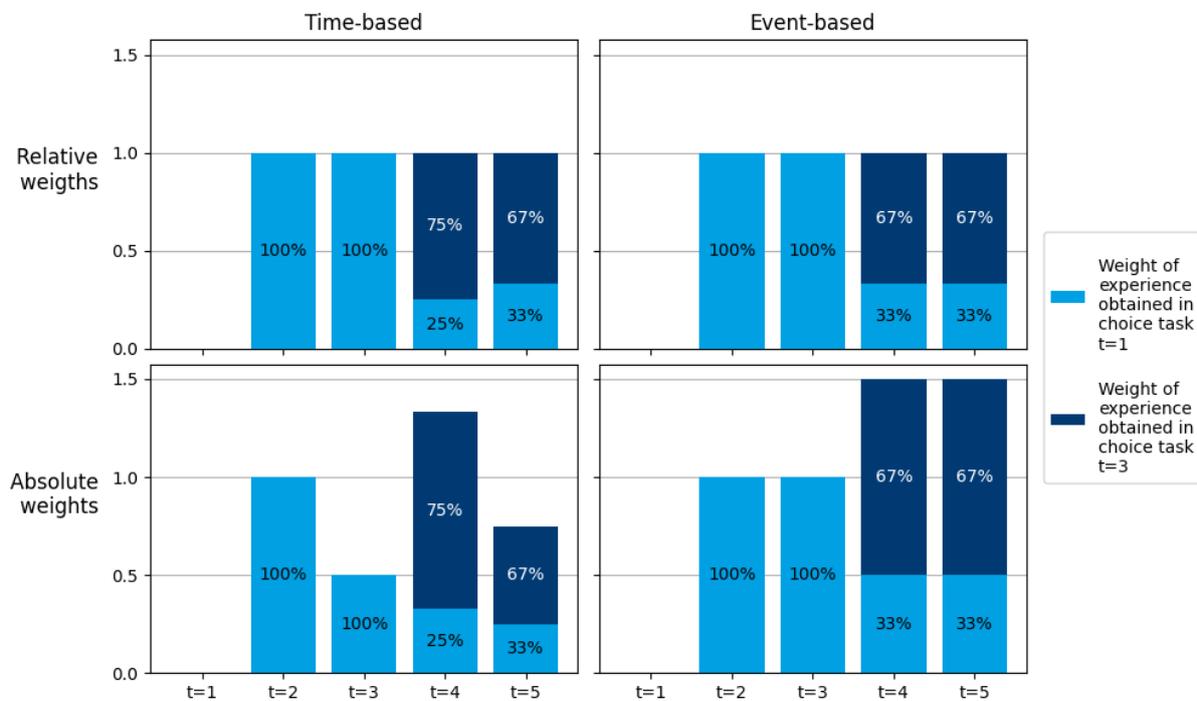

*Figure 2. Weights and their relative importance at a specific choice situation.*
*This specific alternative was chosen at instances t=1 and t=3, with an assumption of d=1.*

Two further experience-related attributes are specified and tested in the model estimation. Firstly, we add two simple binary variables (one per alternative) that mimic the "barrier-to-entry" for using new products and services. Both variables have a value of 1 at the start, as the respondent has no experience with either company. When a company is chosen for the first time, this variable changes to 0 and remains so for the remainder of the scenarios. This allows us to analyse a potential perceived barrier-to-entry that people experience before trying out a service for the first time.

The second experience-based attribute we include is inertia. By incorporating it into the utility function, the inertia parameter captures the value of not having to engage in a serious mental process of decision-making (Cherchi & Manca, 2011). Cherchi & Manca (2011) summarised and tested eight different specifications of choice inertia, from a simple binary specification, to using the performance (attribute level) of the chosen alternatives in previous experiences. Given that the performance of past experiences is already captured through the experiential variables paired with a decay function, we do not test those further. We focus instead on the specifications considering what the previous choices were. These are specifications 1 and 2 according to Cherchi & Manca (2011), which we refer to as "one" and "sum" respectively. "One" is a binary variable, taking the value of 1 if the alternative was chosen in the previous instance and 0 otherwise. "Sum" counts how many times an alternative has been chosen up till that point. To that end, we propose and test three more specifications. Firstly, we formulate a combination of the "one" and "sum" specifications, which we refer to as "reset". Like the "sum" specification, it keeps increasing with the number of times the same alternative is chosen repeatedly. Once the decision-maker switches to a different alternative, the inertia value resets to 0, similar as in the "one" specification. The other two specifications we test are natural logarithms of the "sum" and "reset" specifications. This is to test if the marginal contribution of inertia decreases with an increasing number of repeated choices. For numerical reasons, we add one before applying the logarithm, as a logarithm of zero cannot be defined. Together, the five specifications can be defined as shown in Equation 9. To highlight how the five specifications differ in scale, we plot them in Figure 3. The x-axis indicates which alternative is chosen at a given instance. Note that the inertia is based on a past choice and therefore lags by one choice task. The y-axis shows the value of different inertia variables for alternative A.

*Equation 9. Specifications of choice inertia*

One: $$i_{j,t} = \begin{cases} 0, & q_{j,t-1} = 0 \\ 1, & q_{j,t-1} = 1 \end{cases}$$

Sum: $$i_{j,t} = \begin{cases} I_{j,t-1}, & q_{j,t-1} = 0 \\ I_{j,t-1} + 1, & q_{j,t-1} = 1 \end{cases}$$

Reset: $$i_{j,t} = \begin{cases} 0, & q_{j,t-1} = 0 \\ i_{j,t-1} + 1, & q_{j,t-1} = 1 \end{cases}$$

Log-Sum: $$i_{j,t} = \ln(i(sum)_{j,t} + 1)$$

Log-Reset: $$i_{j,t} = \ln(i(reset)_{j,t} + 1)$$

Where:
- $i_{j,t}$      Choice inertia of alternative *j* in choice situation *t*
- $i_{j,t-1}$      Choice inertia of alternative *j* in choice situation *t-1*
- $q_{t-1}$      Binary variable if alternative *j* was chosen in choice situation *t-1*

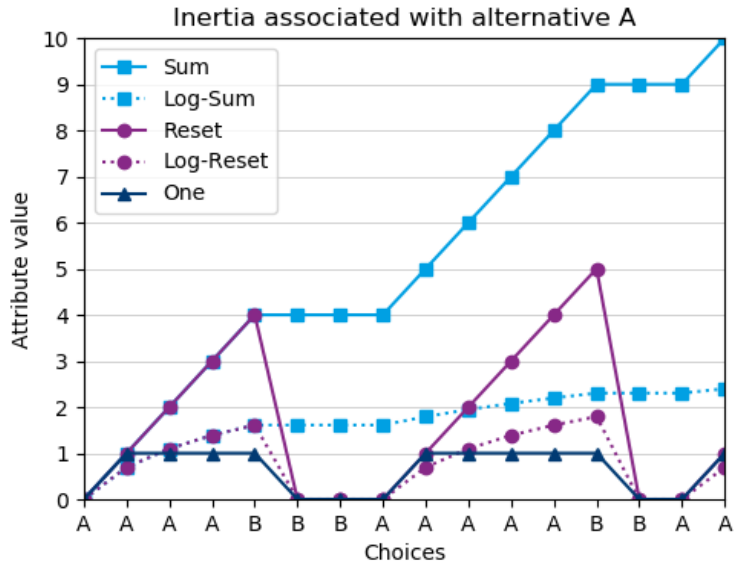

*Figure 3. The contribution of the five different inertia specifications*

## 2.3   Data collection

The full survey was administered to members of the Dutch Mobility Panel (MPN) (Hoogendoorn-Lanser et al., 2015), with the responses gathered between 17-29 of August, 2021. At the time of data collection there were limited regulations regarding the covid-19 pandemic in the Netherlands, with manageable case and hospitalisation numbers and a sizeable part of the adult population being vaccinated (Rijksoverheid, n.d.). Respondents were reimbursed with a flat rate for completing the survey, rather than based on their performance (Ben-Elia et al., 2013), to avoid them focusing exclusively on their financial gain.

In total, 1,304 respondents filled-in the survey, of which 1,023 responses were complete. The data is cleaned based on a minimal and maximal response time. Removing all responses below a certain answering time is standard practice, to remove respondents who rush through the survey without actually reading and considering the task. We set this lower boundary at three minutes. Given the specific nature of our study, a maximum response time is set as well. The MPN allows respondents to stop at any point during the survey and continue at a later time from where they left off. In our experiment, the memory of previous experiences is crucial and by stopping midway and continuing for example a day later, respondents will likely not remember much of what they had seen. We therefore set the maximum cut-off time to 30 min. This reduced the number of observations to 936. We do not apply any answer-based validation, as it might be perfectly possible for a respondent to choose one or the other alternative for all 32 choice sets. The 936 responses are fairly evenly distributed among the nine blocks, with the blocks receiving between 88 and 111 observations.

Socio-demographics of the 936 valid responses are presented in Table 2. The sample is well representative of the Dutch population on most of the presented characteristics. We do observe a slight overrepresentation of older individuals, single-person households and non-urban individuals. The differences in household income can be explained with the respondents in our survey having the option not to disclose their income. Adjusting for that, we see a slight overrepresentation of middle-income households and an underrepresentation of high-income households.

Table 2. Socio-demographic characteristics of the collected sample and the Dutch population. Source for population data: Centraal Bureau voor de Statistiek (2022)

| Variable | Level | Sample | Dutch Population |
|---|---|---|---|
| **Gender** | Female | 52% | 50% |
| | Male | 48% | 50% |
| **Age** | 18-34 | 22% | 27% |
| | 35-49 | 25% | 23% |
| | 50-64 | 29% | 26% |
| | 65+ | 24% | 25% |
| **Education** [1] | Low | 25% | 29% |
| | Middle | 41% | 36% |
| | High | 34% | 35% |
| **Gross household income** [2] | Below average | 20% | 24% |
| | Average | 48% | 47% |
| | Above average | 17% | 29% |
| | Prefer not to say / Don't know | 15% | 0% |
| **Employment status** | Employed | 55% | 50% |
| | Retired | 22% | 21% |
| | In education | 5% | 5% |
| | Other [3] | 18% | 24% |
| **Household composition** | Single person | 26% | 18% |
| | Two or more (with kids) | 40% | 50% |
| | Two or more (no kids) | 33% | 32% |
| **Urbanisation level** [4] | Very highly urban | 23% | 24% |
| | Highly urban | 29% | 25% |
| | Moderately urban | 16% | 17% |
| | Low urban | 23% | 17% |
| | Not urban | 9% | 17% |

[1] Low: no education, elementary education or incomplete secondary education
Middle: complete secondary education and vocational education
High: bachelor's or master's degree from a research university or university of applied sciences

[2] Below average: below modal income (< €29,500)
Average: 1-2x modal income (€29,500 – €73,000)
Above average: Above 2x modal income (> €73,000)

[3] Includes unemployed, unable to work, stay-at-home, volunteer and unknown

[4] Very highly urban: > 2,500 inhabitants/km$^2$
Highly urban: 1,500 – 2,500 inhabitants/km$^2$
Moderately urban: 1,000 – 1,5000 inhabitants/km$^2$
Low urban: 500 – 1,000 inhabitants/km$^2$
Not urban: < 500 inhabitants/km$^2$

# 3 Results

The model estimation outcomes, including both the model fit and the parameter estimates, are presented in Table 3. In addition, Willingness-to-Pay (WtP) trade-offs are presented in Table 4. The presented model achieved a model fit of almost 0.6, with all 13 parameters being highly significant (p<0.01). Four parameters (both waiting time parameters, cost and cancelled service) are specified as random parameters in the ML model, allowing us to capture how their perception varies within the population. The remaining five parameters are estimated as fixed.

*Table 3. Model outcomes*

| | | |
|---|---|---|
| **Parameters** | 13 | |
| **Null LL** | -20,761.14 | |
| **Final LL** | -8,370.43 | |
| **Adjusted Rho-square** | 0.596 | |
| **BIC** | 16829.8 | |

| | Parameter | Robust t-stat [parameter] | σ | Robust t-stat [σ] |
|---|---|---|---|---|
| **Constant** [*company B*] | -1.220 | *-5.30* | | |
| **Trip cost** | -0.625 | *-5.21* | 0.867 | *13.70* |
| **Waiting time** [*early pick-up*] | -0.277 | *-7.86* | 0.289 | *6.00* |
| **Waiting time** [*late pick-up*] | -0.189 | *-21.60* | -0.173 | *-11.30* |
| **Cancelled ride** | -2.780 | *-27.30* | 1.500 | *11.30* |
| **Barrier-to-Entry** | -1.240 | *-12.50* | | |
| **Inertia** | 0.058 | *9.14* | | |
| **Memory decay parameter** [*waiting time*] | 1.670 | *-8.76* | | |
| **Memory decay parameter** [*cancelled service*] | 1.660 | *-10.40* | | |

Waiting time is specified as the difference between the expected and experienced waiting times, with a positive value indicating a late pick-up (waiting longer than expected) and a negative value representing an early pick-up. A separate parameter for estimated waiting time is tested, but resulted in a minimal improvement in model fit (five log-likelihood points), so it is not considered further. By specifying separate parameters for early and late pick-ups, we can see a significant difference in their perception. The results indicate that a minute of saved waiting time (earlier than expected) is equally as positive as 1.5 minutes of longer-than-expected waiting time is negative. In terms of monetary trade-off (Table 4), travellers are willing to pay €0.44 for each minute of saved travel time if the pick-up is realised earlier than planned, or €0.30 for each minute if it is later. This is somewhat counterintuitive, as a higher penalty (and thus higher WtP) would be expected for longer-than-expected waiting times. We discuss this further in the following section. Both parameters are modelled as random, and we can see from Figure 4 that although waiting times with an early pick-up are perceived more negatively, their perception is also subject to greater variability, whereas a longer-than-expected waiting time is more consistent across the sample.

If the ride is cancelled by the driver or platform, after the traveller has already made their decision, this results in quite a strong penalty, equal to a price increase of €4.45. Or in other words, this is the discount needed in the following travel instance for the traveller to consider this company for their travel choice.

Barrier-to-Entry is added as an attribute to mimic the initial hesitation of trying a new product or service. Our model estimates show that this barrier is equal to approximately €2.00, meaning that such a discount may prove sufficient to entice users to try this new service provider.

To test for the impact of socio-demographics (income, car ownership, frequency of car use), service familiarity and situation perception (Table 6 in the Appendix), several models with interaction effects are also specified. As no interaction yielded a significant improvement in model fit and, in most cases, also no significant parameter estimates, interactions are not included in the final model specification.

| *Table 4. Willingness-to-Pay for aspects of the trip* | |
|---|---|
| **Attribute** | **WtP** |
| Waiting time with early pick-up | 26.59 €/h |
| Waiting time with of late pick-up | 18.14 €/h |
| Cancelled ride | 4.45 € |
| Barrier-to-Entry | 1.98 € |
| Preference for company B | 1.95 € |
| Marginal value of inertia* | 0.09 € |

*per additional experience*

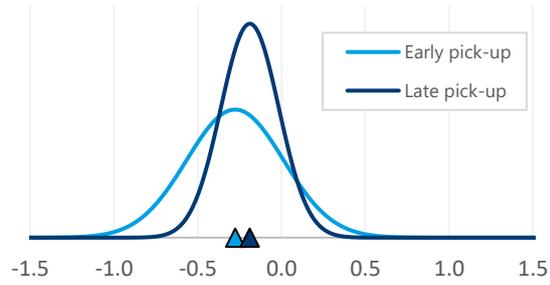

*Figure 4. Distribution of waiting time parameters*

Turning to memory decay and how it is captured, we test four different approaches, as detailed in Section 2.2. In Table 5, we summarise the outcomes for the four different specifications[1]. We observe that an event-based formulation with absolute weights seems to perform best for our sample, followed closely by the time-based relative weights formulation, as specified by Tang et al. (2017).

Given the best-performing model specification, the power decay function estimated in the model is presented in Figure 5. Although we estimate two separate decay functions, one for waiting time and one for a cancelled ride, the estimates are nearly identical, so this figure shows a single line for clarity. What is striking is that the relevance of past experiences seems to decay rapidly, with the second instance already having only a third of the impact of the latest. The fourth instance has a weight of ~0.1 and by the 7th, the weight drops below 0.05. This seems to indicate that respondents only kept a handful of the most recent instances in mind and discarded the rest, with the most recent experience weighing by far the most. As this is the absolute-weight formulation, it is interesting to consider the cumulative weight of all past experiences except for the first one. Given the sharp decline in importance, a large number of experiences would be required to cumulatively weigh more than only the first. Figure 5 shows that nine experiences (2-10) give a combined weight of 0.8 (44% of the total weight), compared to the most recent instance (always taking a value of 1). And even considering 31 instances (the maximum in our experiment), the combined weight of all experiences bar the first sums up to 0.967.

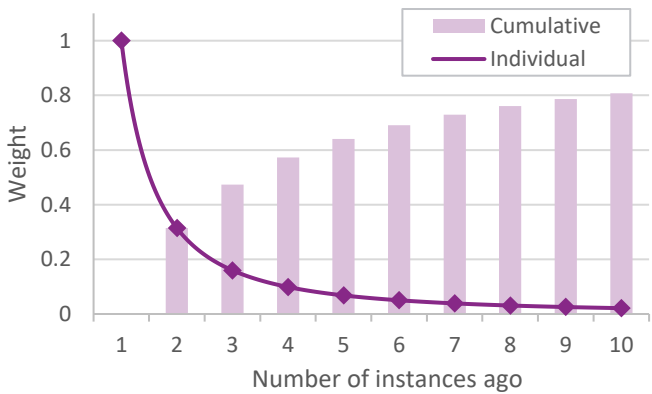

*Figure 5. Power decay function*

*Table 5. Model fits (log-likelihood and adjusted rho-squared) of different memory decay specifications*

|  | **Time-based** | **Event-based** |
|---|---|---|
| **Relative weights** | -8,776.15 *(0.5914)* | -11,581.78 *(0.4417)* |
| **Absolute weights** | -9,920.75 *(0.5217)* | -8,445.46 *(0.5927)* |

Lastly, we comment on the various inertia formulations. The best performing on our dataset is the reset specifications, where the variable increases by a value of one each time the same option is chosen, then drops to 0 once the respondent switches to a different alternative. This is followed by the log-sum and log-reset specifications, although the reset specification is significantly better performing. Table 4 shows that each additional experience adds a value of €0.09. In other words, said alternative could get €0.09 more expensive after each instance and the respondent would still not switch to a different alternative (ceteris paribus). It also means that a potential competing company would need to offer a discount of the same magnitude for each additional experience the traveller has with their rival.

---

[1] The model fit in Table 5 differs slightly from the model fit reported in Table 3, as the specification of memory decay (absolute and relative weights; time- and event-based formulation) was tested first. Once the best approach was obtained, different specifications of taste parameters were tested.

## 4 Conclusion and Discussion

This paper presents novel insights into the behaviour of individuals in response to unexpected outcomes and how such experiences impact future decision-making. We perform a stated preference experiment on the topic of ride-hailing waiting time on a representative sample of the Dutch population. Mixed logit models are used to quantify the valuation of reliability and capture the heterogeneity within the sample.

Our findings show that unexpected waiting time is valued at approximately 20-25€/h. This is somewhat higher than the roughly 15€/h which was uncovered by Alonso-González et al. (2020). Both studies are based on SP data obtained in the Netherlands, although the works differ in their approach for studying reliability, which is where the difference may originate from. Alonso-González et al. (2020) provided reliability information explicitly (upfront), whereas in our survey, respondents learned the reliability of a service implicitly, through trial-and-error (as explained in more detail in Section 0). The only other study to have considered waiting time reliability of ride-hailing (or other forms of on-demand mobility) does not allow for WtP comparison, as no cost component was included in the survey (Bansal et al., 2019). Comparing our findings by means of a ratio between waiting time and travel time is not feasible. To limit the cognitive burden for respondents, we did not vary travel time, meaning we cannot estimate it. Typically, waiting time tends to be perceived 2-3 times more negatively than in-vehicle time (Wardman, 2004). Considering the approximate value-of-time for the Netherlands of 10 €/h, as reported by de Jong & Kouwenhoven (2019), our findings seem to be well within the expected range.

Considering the different valuation of early (26.59 €/h) and late pick-ups (18.14 €/h), our results may seem somewhat counterintuitive. We expected the deviation from the estimated waiting time to be perceived as a dissatisfier (Van Hagen & Bron, 2014), meaning that travellers expect the promised level of performance. There is no benefit associated with performing as expected (on-time) or better (early), but it results in disutility if the performance is below expectations (late). This was also tested by means of a random regret minimisation model (Van Cranenburgh et al., 2015), which is able to discern such effects, but was not found for our dataset. A possible explanation is that waiting time is not seen as a dissatisfier, and that travellers are (very) positively surprised by an early arrival which may make the overall experience more enjoyable. This latter interpretation means that service providers should always strive to be as early as possible, regardless of the displayed waiting time. This however may be troublesome for multiple reasons. As passengers know that the vehicle will wait for them (contrary to traditional public transportation services), some may not arrive at the pick-up location earlier than the displayed departure time, resulting in the vehicle and therefore driver and potential co-riders needing to wait, making their experience less enjoyable (Kucharski et al., 2020). It is also fairly easy to manipulate the upfront information, with service providers displaying a very "safe" expected waiting time and then regularly performing better than expected. However, Alonso-González et al. (2020) indicate that waiting time variability is perceived less negatively than the initially displayed waiting time, meaning that a balance between low anticipated waiting time and striving to achieve it is required. Further experiments, possibly enriched by RP data, will hopefully shed more light onto this topic.

The valuation of cancelled services has, to the best of our knowledge, not yet been investigated, making is thus difficult to validate. A value of €4.45 or 10-15 waiting-time-equivalent minutes may seem on the low end, especially considering that in our hypothetical case, a cancellation notice is only given after the estimated waiting time has elapsed. This somewhat low value may be a consequence of the SP nature of our data, as respondents did not truly experience the downsides of the cancellation. Another potential cause may be the attribute levels of cost used in the experiment (varied between €12.00 and €15.00). However, parallels can be drawn with the perception of denied boarding in public transport. Yap & Cats (2021) looked at the perception of waiting time pre and post denied boarding and found them to be 1.6 and 2.7 times the value of in-vehicle time respectively. Calculating the fixed penalty from the relative waiting time penalty reported by Yap & Cats (2021) shows that our service cancellation would be on the high end, equating to a waiting time of 20 to 25 minutes[2].

With respect to the barrier-to-entry estimate, as it was not directly introduced in the survey, it can be seen only as a proxy for such an effect. That also means that comparison with other values is difficult. However, the parameter is (highly) significant and approximate value does point to a potentially necessary discount to entice new users to try a service, which is often similar to what many new market entrants utilise to generate demand (trial periods with a discount, first-time use free,…).

A major contribution of this paper is also testing different memory decay specifications as well as showcasing the relative importance of past experiences. The power function parameter obtained a fairly high value of 1.67 and 1.66 for the waiting time and cancelled service respectively. As is highlighted in Figure 5, this means that the importance of past instances decreases quickly, but it also means that comparatively, the most recent experience carries a lot of weight. Service providers are therefore required to offer significant concessions to customers after a particularly bad experience if they wish to retain them. That bad experience will fade quickly however, once newer (better) experiences are gathered. Albeit in a different context, Tang et al. (2017) report a very similar value of the decay parameter, namely 1.64, in their "uninformed" group of respondents. They split the respondents into "informed" (receiving real-time updates throughout the trip) and "uninformed" (only receiving static information before the trip). The "uninformed" group is therefore essentially identical to our experiment, showcasing that in a similar setup, the context does not seem to have a significant impact on the rate of memory decay. It is interesting to observe that Tang et al. (2017) report a much lower value of only 1.11 for the "informed" group, meaning that the last experience carries much less weight and a larger set of experiences are likely considered when making a choice if the respondent is kept up-to-date when travelling.

It is also interesting to compare the four different memory decay specification, mainly with the two that performed far better. The fact that they are on the diagonal in Figure 2 indicates that they are the most different from each other and thus capture distinctive aspects of memory decay in different ways. The time-based relative-weights approach has continuously decreasing weight value with each new instance, but due to the relative nature of the weights, only the ratio between them changes, which is logical as when experiences are further away, it is less relevant if it happened say 14 or 15 days ago, as opposed to 1 or 2 days ago. On the other hand, this approach does not allow for differentiating between a different number of experiences (be it good or bad). Considering an individual had 10 equally bad experiences with one company and only 1 bad experience of equal magnitude with another, the averaging of weights means that the overall disutility contribution is equal. This is what the absolute-weight approach solves, but by staying with a time-based memory specification would mean that the sum of all weights can drop below 1. This is solved by the event-based approach, where there is always an instance that is valued with a weight of 1. The downside of this approach is that over time, the ratio between experiences does not change. Even though our results indicate a superior performance of an event-based absolute-weight formulation, this may very well be experiment-specific. The nature of SP experiments also makes it difficult to make a strong differentiation between what we characterise as time-based or event-based memory. Given that in our survey respondents answered all 32 scenarios in one go, it is difficult to say for certain which specification is better. Utilising RP data, where the passage of time is much more evident and also clearly recorded may provide relevant further insight into this topic and the role of elapsed time.

A point of attention in IBL, that is present in both specifications, but may be particularly prominent in the absolute-weight formulation, is serial correlation. IBL uses past instances to predict future choices of respondents, and information on past instances is only recorded for the selected alternatives. This is then exacerbated in the absolute-weights approach, as the contributions of past experience accumulate for a single alternative, meaning the utility of said alternative will grow in magnitude and a substantially larger offset (bad experience) will be required for another alternative to become appealing. This is less of a concern in the relative-weight scenario, as all the weights for all alternatives always equal one and only the magnitude of the experiences may differ.

---

[2] Depending on the waiting time the users would experience after being denied boarding, the penalty would be roughly between €1 and €3 for waiting times of 5 min and 15 min respectively.

The specification of memory decay likely also has a strong influence on the performance of different inertia (habit) specifications. The additive nature of the absolute-weights approach means that by continuously selecting the same option, more weight is put onto it, resulting in an ever larger (dis)utility. The reset specification is able to offset this additional (dis)utility. It is striking that all three of the specifications developed in this research outperformed the "one" and "sum" specifications. We do note that Cherchi & Manca (2011) also test specifications utilising the actual performance of alternatives, which we do not test, as those are indirectly included through the memory decay function. As with the memory decay, the inertia specification may also be influenced by the SP nature of the experiment and thus data based on revealed behaviour may result in a different specification performing best.

These insights – both the taste related findings on time valuation, ride cancellation and barrier-to-entry, as well as the memory decay and inertia workings of decision-makers – are potentially valuable insights for policymakers, authorities and operators for designing, governing and regulating on-demand services. When providing travellers with a waiting (and also travel) time estimates, a balance between the descriptive (upfront displayed waiting time) and experiential (actual waiting time) information is necessary. Analysing the barrier-to-entry shows the advantage that the first entrants to the market will have, as they do not have to entice users from a competing service to try out their own. However, their position may be even more difficult, as a switch between different providers of the same type of service may be easier for individuals than trying a completely new service in the first place. Here, existing providers of a different service (in particular public transport operators) could have the upper hand, being more familiar to the local population, or at the very least by its existing users, and thus may find it easier to launch a new type of service within its existing portfolio. Finally, in order to retain users, it is key for the operators to offset bad experience immediately after they occur, ideally through financial incentives, as their impact on other relevant attributes (travel and waiting time) is limited. The upside is that if carried out correctly, a good experience will quickly balance out the bad and operators do not have to suffer the consequences of a single bad instance for a long time.

## Acknowledgement

This research was supported by the CriticalMaaS project (grant number 804469), which is financed by the European Research Council and Amsterdam Institute for Advanced Metropolitan Solutions.

## Conflict of interest

On behalf of all authors, the corresponding author states that there is no conflict of interest.

## CRediT author statement

**Nejc Geržinič:** Conceptualization, Methodology, Software, Validation, Formal analysis, Investigation, Data Curation, Writing – Original Draft, Visualization **Oded Cats:** Conceptualization, Methodology, Writing – Review & Editing, Supervision, Project administration, Funding acquisition **Niels van Oort:** Conceptualization, Methodology, Writing – Review & Editing, Supervision **Sascha Hoogendoorn-Lanser:** Investigation, Writing – Review & Editing **Michel Bierlaire:** Methodology, Writing – Review & Editing **Serge Hoogendoorn:** Writing – Review & Editing, Supervision

# Appendix

*Table 6. List of additional questions and possible answers, presented to the respondents*

| | Question | Possible answers |
|---|---|---|
| 1 | In the event of a cancelled services, how would you have gotten home? | - Using the other FLEX company<br>- Taxi<br>- Public transportation<br>- Ask a friend / family member to drive me<br>- Use a shared car or shared bike<br>- Walk<br>- Other, namely… |
| 2 | How would you use FLEX in the future, after getting a ride cancelled? | - I would keep using FLEX without change<br>- I would be more cautious using FLEX, making sure I have alternative means of transportation<br>- I would not use that FLEX company anymore<br>- I would avoid using FLEX as much as possible<br>- I would not use FLEX anymore |
| 3 | Which company did you find more reliable? | - Company A<br>- Company B |
| 4 | Where did you imagine yourself waiting for the ride? | - Sitting inside<br>- Standing inside<br>- Sitting outside<br>- Standing outside<br>- I did not think about it |
| | How familiar are you with… | |
| 5 | …car sharing? | - Never heard of it |
| 6 | …OV fiets? | - Familiar with it, but never used it |
| 7 | …bike / scooter sharing? | - Used it once |
| 8 | …flexible public transport? | - Used it a few times |
| 9 | …ride-hailing? | - Use it regularly |
| 10 | …food delivery services? | |
| 11 | …home rental services? | |